# Climate Anomalies vs Air Pollution: Carbon Emissions & Anomaly Networks[1]

**Anshul Goyal**
Carnegie Mellon University
Pittsburgh, PA, U.S.A
anshulgo@andrew.cmu.edu

**Kartikeya Bhardwaj**
Carnegie Mellon University
Pittsburgh, PA, U.S.A
kbhardwa@andrew.cmu.edu

**Radu Marculescu**
Carnegie Mellon University
Pittsburgh, PA, U.S.A
radum@ cmu.edu

## ABSTRACT

*This project aims to shed light on how man-made carbon emissions are affecting global wind patterns by looking for temporal and geographical correlations between carbon emissions, surface temperatures anomalies, and wind speed anomalies at high altitude. We use a networks-based approach and daily data from 1950 to 2010 [1-3] to model and draw correlations between disparate regions of the globe.*

## 1. INTRODUCTION & MOTIVATION

Earth's climate is changing rapidly and there are good reasons to believe that the release of carbon emissions by human activity is one of the many causes. This climate change comes in many forms such as variations in precipitation patterns and variations in storm patterns. One of the lesser studied forms is the change in wind patterns, both at low and high altitudes.

This project will look at global carbon emissions, global temperature anomalies, and global wind pattern anomalies at high altitude to see if there are geographic and / or temporal correlations between the three metrics. Identifying these correlations can help us understand what is driving certain features of our changing climate such as: (i) Why the polar caps are warming faster than the equator? [6] (ii) How are carbon emissions being distributed in our atmosphere? [6] (iii) Why are common paths for hurricanes changing? (iv) Are these variations correlated to increased carbon emissions in the atmosphere?

## 2. PREVIOUS WORK

Research into changes in Earth's climate is ongoing. It is becoming more and more relevant as people across the globe appear to experience more chaotic weather every year. Much of this research has gone into studying changes in various climate patterns (precipitation, temperature, wind). [8] Of the research that studies wind patterns, some of it documents the disagreement among different climate models when it comes to trends for global wind patterns [7], while other publications draw contradicting conclusions depending on the form of data used. [8] Fortunately, there seems to be some consensus that winds across U.S. (and even North America) are slowing down, but only at the surface. [6 - 8]

## 3. DATA

Daily data from 1951 to 2010 is derived from the Carbon Dioxide Information Analysis Center [2] for global carbon emissions, from the Met Office [1] for global temperature anomalies, and from University Corporation for Atmospheric Research [3] for global wind velocity. Using the raw wind velocity data from [3], we derive daily wind speed anomaly data.[2]

In addition, the data for each metric has been resized to fit a global network of nodes that has a geographic grid size of 2.5° x 3.75° (2.5° between latitudes, 3.75° between longitudes). This gives us a global grid that spans 96 nodes in the East-West direction and 73 nodes in the North-South direction. This resolution is due to the temperature anomalies data from [1]. Data for other metrics was resampled using nearest-neighbors to fit the lower resolution.

At this resolution, our global grid has 7008 nodes (compared to 726 nodes in [4]) and roughly 24.5 million links (fully connected). However, the data for carbon emissions and temperature anomalies is for land surfaces only while data for wind spans across the entire globe (see Figures 1 and 2). This leads our carbon emissions and temperature anomalies networks to have an effective size of 1330 nodes with about 880,000 links and 1619 nodes with about 1.3 million links, respectively.

## 4. APPROACH

For this project, we have borrowed and modified the approach used in [4]. Our approach involves constructing four fully-connected networks for each of the following metrics: carbon emissions, surface temperature anomalies, and wind speed anomalies at the 500 millibar pressure level (an altitude of about 5.5 km). Each network is used to model the globe, with each node representing a certain region of the globe. Each of the four networks has a specific purpose that is discussed in the following sections.

### 4.1 Regular Networks

The main purpose of the first two networks (the "regular" networks) for each metric is to model daily global activity for that metric from 1950 to 2010 using the gathered data. One of the two networks is used to find positive correlations while the other is used to find negative correlations (anti-correlations). We do this by calculating the cross-covariance between every pair of nodes in the network. The values of the cross-covariance are then used to calculate the weight of the link between pairs of nodes using the expressions in Equations 1 and 2.

These expressions show how we derive the weights $\boldsymbol{P}$ and $\boldsymbol{N}$ between nodes $\boldsymbol{m}$ and $\boldsymbol{n}$ for year $\boldsymbol{y}$ using the cross-covariance values $\boldsymbol{C}$. The regular network used to find positive correlations has link weights defined by $\boldsymbol{P}$ while the other

[1] This is a class project report for CMU course 18-755 in Fall 2016.

[2] The anomaly data is calculated for any particular day by taking the value for that day and subtracting the average of the same day over all the years.

regular network (used to find negative correlations) has link weights defined by $\boldsymbol{N}$.

$$P_{m,n}^{y} = \frac{max(C_{m,n}^{y}) - mean(C_{m,n}^{y})}{std(C_{m,n}^{y})}$$

**Equation 1: Positive Covariance Link Weight**

$$N_{m,n}^{y} = \frac{min(C_{m,n}^{y}) - mean(C_{m,n}^{y})}{std(C_{m,n}^{y})}$$

**Equation 2: Negative Covariance Link Weight**

Just like the approach used in [4], many of the link weights defined by $\boldsymbol{P}$ and $\boldsymbol{N}$ will likely be due to auto-correlation. The purpose of the next two networks for each metric is to find the possible link weights due to auto-correlation alone, and then remove those links from the two regular networks. These next two networks are our "surrogate" networks. Of note, we have two surrogate networks for the same reason we have two regular networks: one for positive link weights, one for negative link weights.

### 4.2 Surrogate Networks

To find the maximum possible link weight due to auto-correlation alone, we calculate link weights for the surrogate networks using the expressions in Equation 1 and 2 (the same way we calculated link weights for our two regular networks), but the data for the surrogate networks is temporally shuffled (but not geographically shuffled). As was done in [4], we shuffle the data by choosing a random sequence of years for each node in the network (but keeping the order of the days within each year intact). This is done to preserve auto-correlation and node statistics but break the physical dependence between nodes.

Using the weight of the links in the surrogate networks, we define a threshold for each regular network by visually estimating the maximum weight of the links in the corresponding surrogate network. We use visual estimation rather than the maximum link weight of the surrogate network because the order of the year chosen as our random sequence may still cause spurious correlations that lead to a few high link weights. Using this threshold, any links in the regular network which have a weight greater than the threshold will be considered real (i.e. not due to auto-correlation) and are preserved. Links with a lesser weight are removed. See Figures 3 to 8 for a distribution of the link weights for both regular and surrogate networks across all years for all three metrics.

As an example, let's look at the link weight distribution for the temperature anomalies network with positive covariance. We can see in Figure 3 that the regular network's ("real") link weight distribution spans as high as 6, while the surrogate network's ("shuffled") link weight distribution only spans as high as 4. Using this information, we remove links in our regular network that have a weight that is less than or equal to 4.

With weaker links removed from the regular networks of each metric, there are still thousands of links in the network. To focus our analysis, we define an even higher threshold to remove the weakest links and preserve only the 200, 100, or 50 heaviest links in the network. While the links we remove at this point are likely modeling real phenomena, we chose to look at the heaviest links to identify a pattern in our networks with the highest likelihood of modeling a real change in the climate.

With more time, we would have attempted to find geographic and temporal correlations between the highly connected nodes between networks. We can do this by matching up the nodes for each network (by longitude and latitude) and then running multi-network community detection. The hope is to find communities of nodes that span across the networks. This would signify areas of the globe that experience anomalous behavior at the same time across different metrics (i.e. wind anomalies, temperature anomalies, carbon emissions).

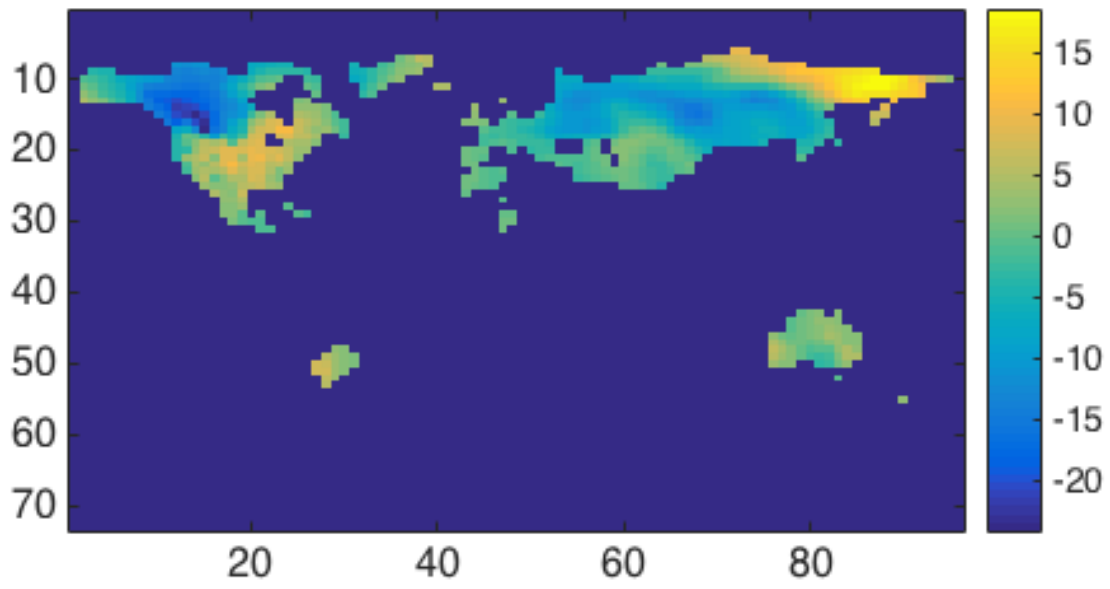

**Figure 1: Anomalous Temperature on January 1st, 1950**

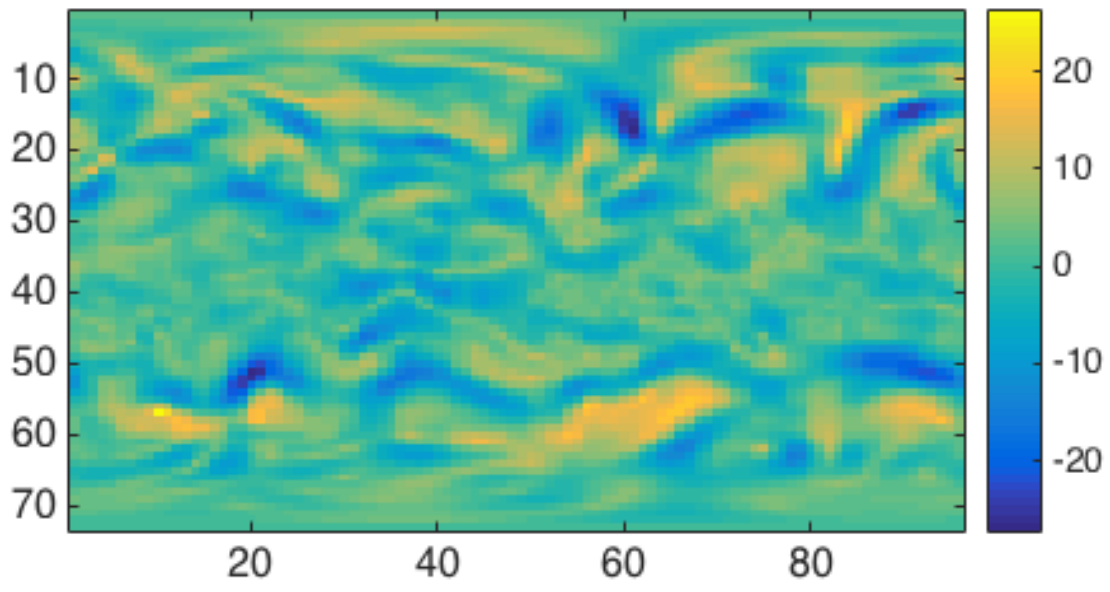

**Figure 2: Anomalous Wind at 500 Millibar on January 1st, 1950**

### 4.3 Scalability

Even though we have borrowed almost our entire approach from the approach used in [4], there are a couple of key differences. First, our approach is a scaled-up version that includes multiple networks, each of which is a higher resolution global network than the global network used in [4] (see Data section above). The hope is that a higher resolution network may lead to more interesting results, but a larger number of nodes does require a significant increase in computational resources and computing time to handle calculating link weights between ever pair of nodes.

Second, the approach used in [4] effectively treated positive covariance and negative covariance as the same by taking the absolute value of the cross-covariance between pairs of nodes. This could potentially eliminate significant links between nodes that observe both a large positive and large negative covariance, at different time delays. In our approach, we are generating separate regular and surrogate networks for positive and negative cross-covariance to preserve these significant links, if they exist.

## 5. EXPERIMENTAL SETUP

We used MatLab for all data calculations and manipulations for all three metrics: carbon emissions, surface temperature anomalies, and wind speed anomalies at the 500 millibar pressure level. To provide some context, see Figures 1 and 2 for the data visualization for January 1st, 1950 for anomalous temperature and anomalous wind at 500 millibars.

In addition, we have calculated positive covariance link weights and negative covariance link weights for regular and surrogate networks for temperature anomalies and carbon emissions, plotted the histograms for each pair of regular and surrogate network link weights (see Figures 3 to 8), the number of links in the resulting networks (see Figures 9 to 14), and visualizations for the resulting networks (see Figures 15 to 16).

## 6. RESULTS

In this section, we present the results of our approach. We start with the distribution of link weights for each network to justify our selected threshold weights and special treatment of the threshold for our wind anomalies networks. We then focus on the heaviest links in the wind anomalies network.

Figures 3 to 8 plot the distribution of all positive and negative covariance link weights for our regular and surrogate networks for temperature anomalies, carbon emissions, and wind speed anomalies. These histogram plots show us if there are links between nodes due to positive cross-covariance and negative cross-covariance that are likely modeling real phenomenon.

For the temperature anomalies networks, using the surrogate network link distribution (Figures 3 and 4), we chose a link weight threshold of 4. Above this threshold, the regular networks contain links that are likely modeling temperature anomaly correlations between regions of the globe due to real phenomenon. Similarly, for the carbon emissions networks, using the surrogate network link distribution (Figures 5 and 6), we chose a link weight threshold of 3.

Figures 9 and 10 show the number of links per year for each carbon emissions networks. Figure 9 appears to show an increasing trend over time for positively covariant carbon emissions. This intuitively makes sense since humanity is increasing fossil fuel consumption over time and, therefore, releasing more carbon emissions over time. In comparison, Figure 10 shows a low number of links for most years except for 1986, 2002, and 2005. As of now, we are not sure what caused these spikes but this is an area of possible future investigation.

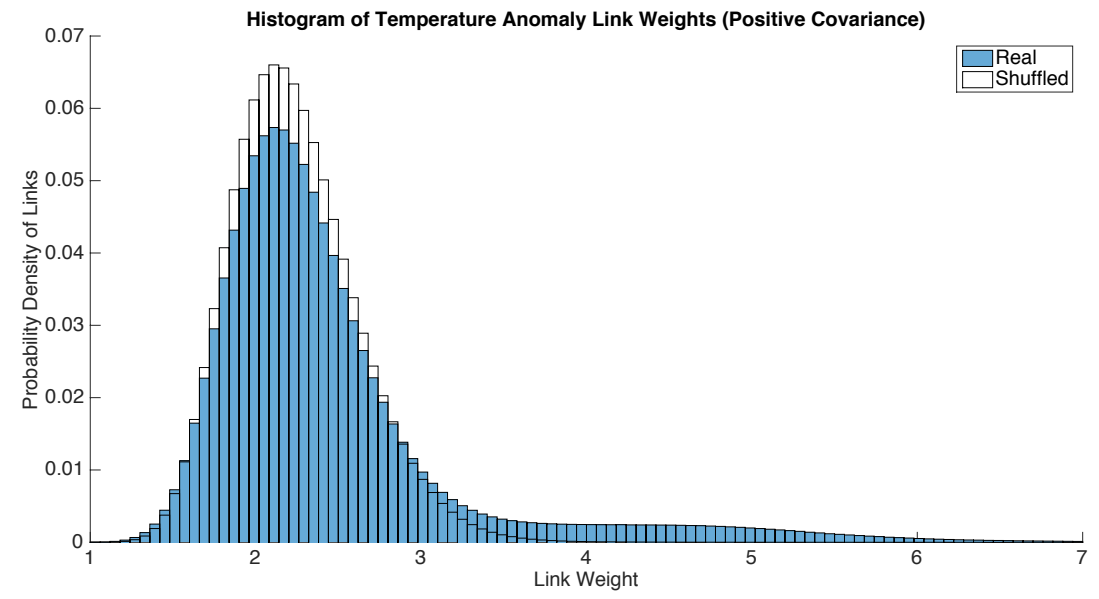

**Figure 3: Histogram of Temperature Anomaly Link Weights with Positive Covariance**

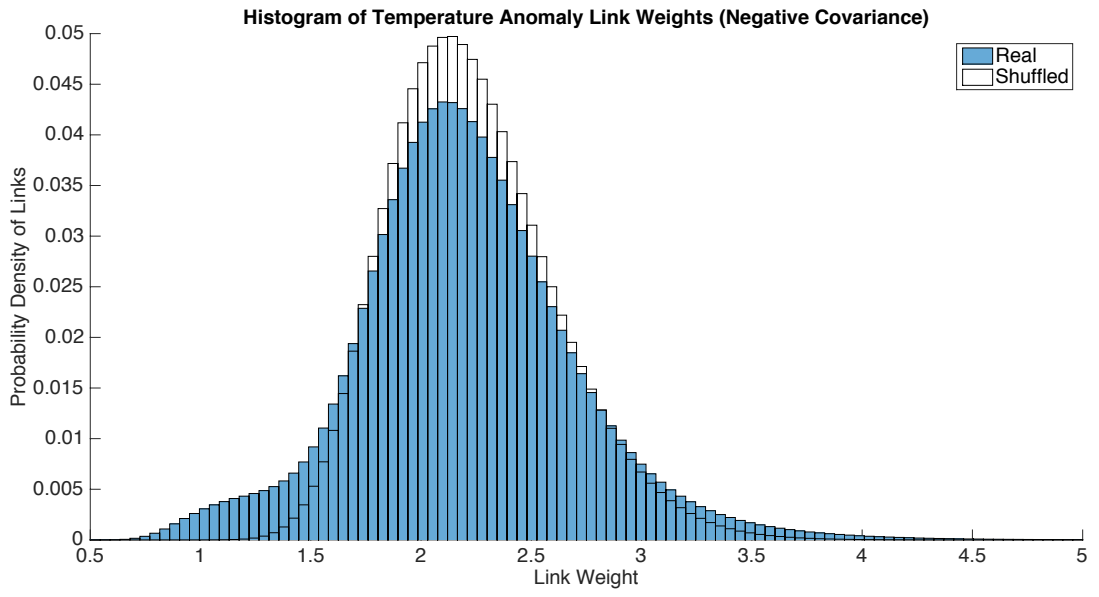

**Figure 4: Histogram of Temperature Anomaly Link Weights with Negative Covariance**

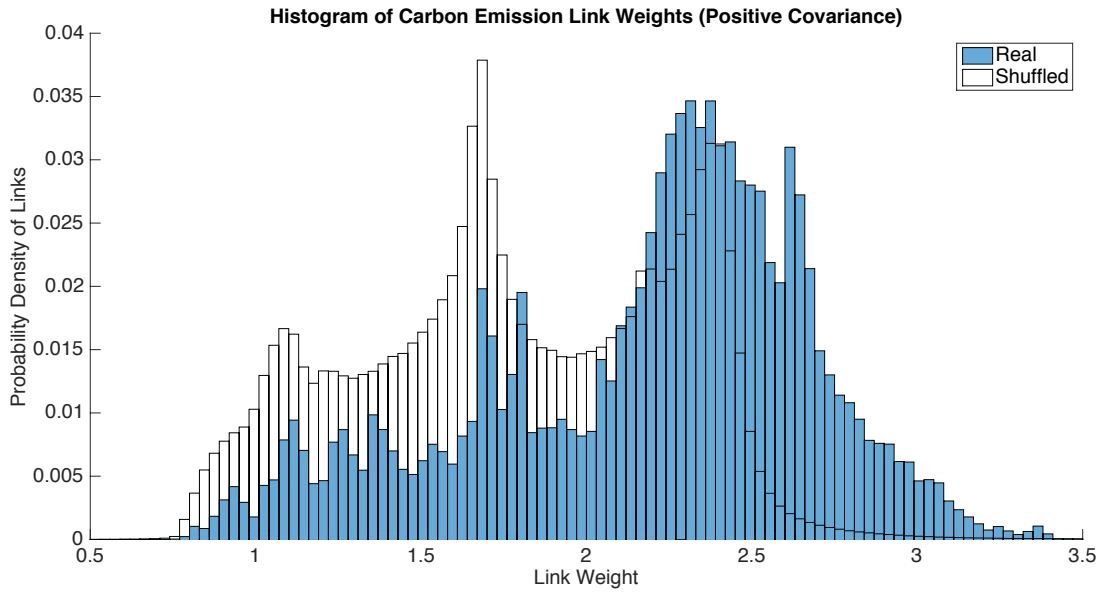

**Figure 5: Histogram of Carbon Emissions Link Weights with Positive Covariance**

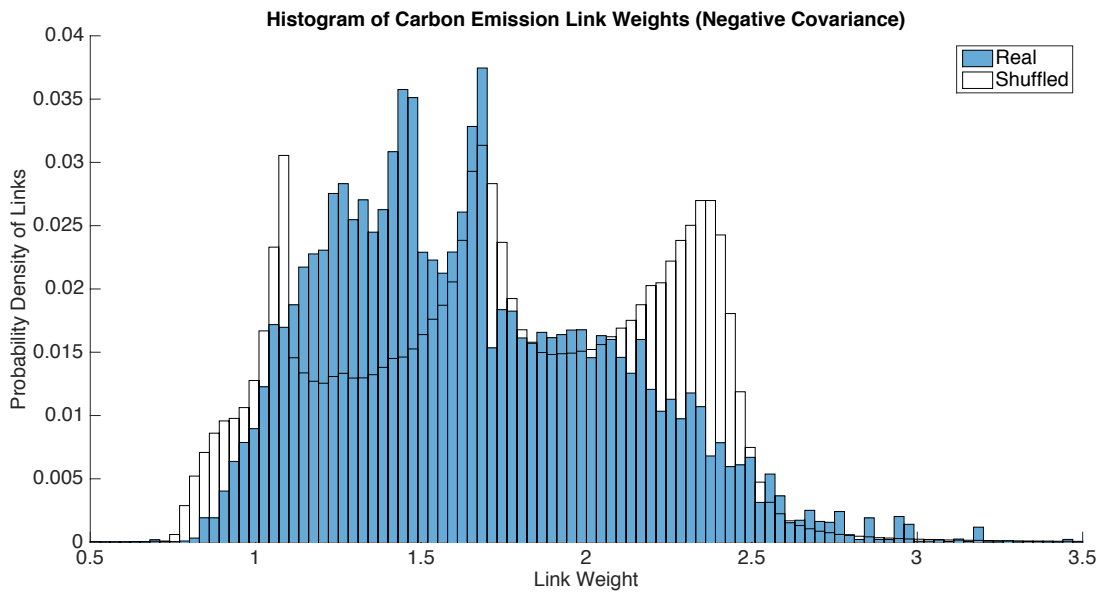

**Figure 6: Histogram of Carbon Emissions Link Weights with Negative Covariance**

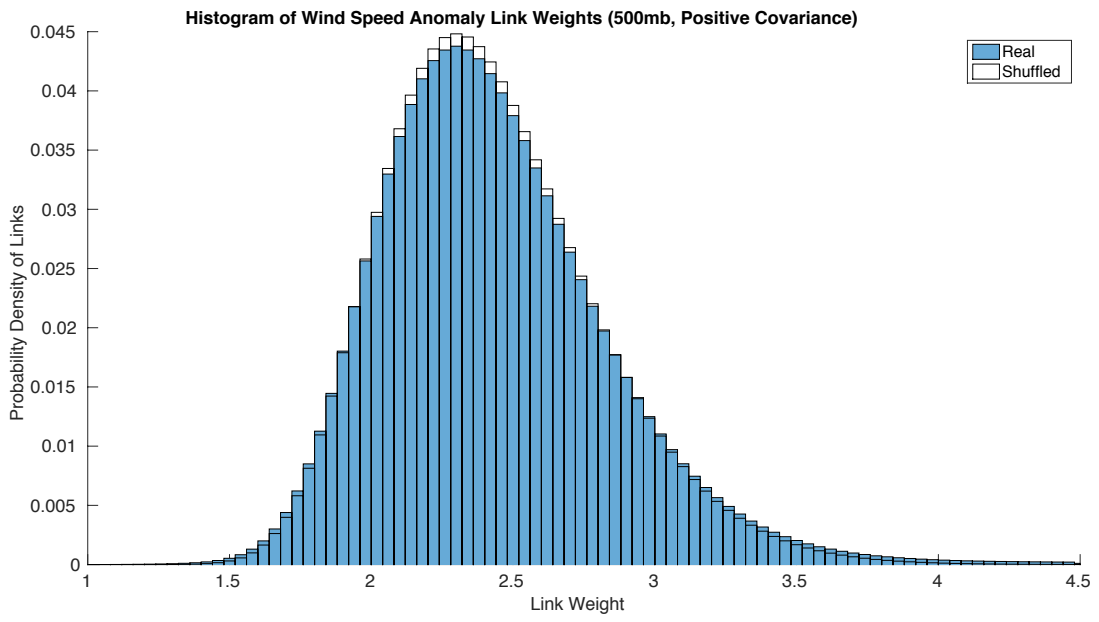

**Figure 7: Histogram of Wind Speed Anomaly Links Weights with Positive Covariance**

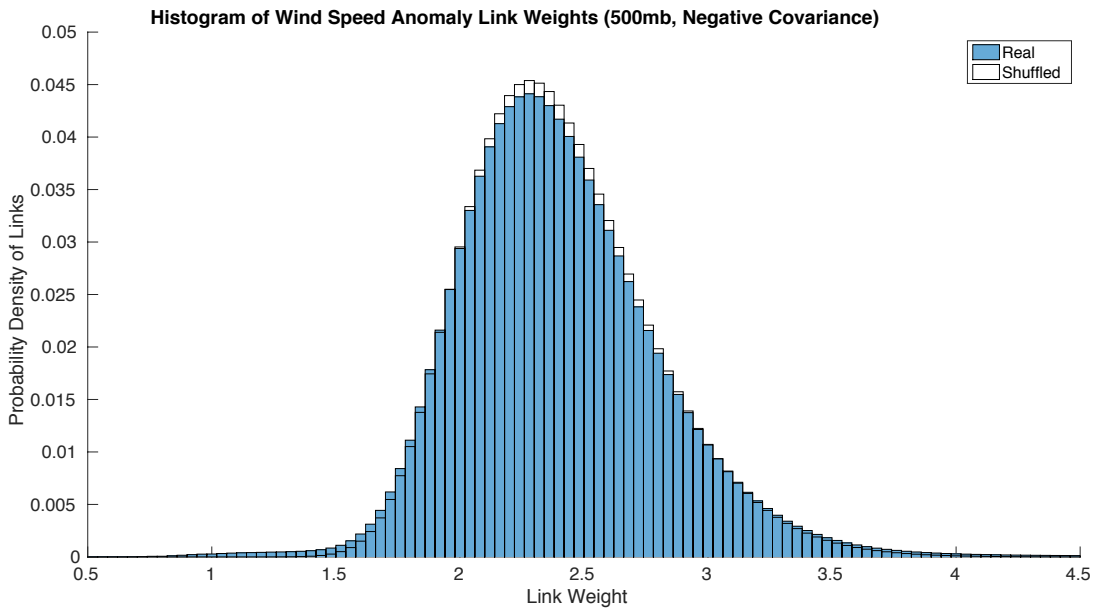

**Figure 8: Histogram of Wind Speed Anomaly Links Weights with Negative Covariance**

For the wind speed anomalies networks, the link weight distributions for our regular and surrogate networks are so similar that it is difficult to find a clear threshold (see Figures 7 and 8). To make sure we chose a threshold that would remove all possible auto-correlated links, we decided to use the maximum link weight of the surrogate networks to define the threshold for their respective regular networks (8.89 for our positive covariance network and 6.34 for our negative covariance network). Even though these threshold values are high (compared to the threshold for our temperature anomalies and carbon emissions networks), the regular networks still contained many links after thresholding out weaker links (23,417 links for our positive covariance network and 180,685 links for our negative covariance network).

Figures 11 and 12 show the number of links per year for each of our wind speed anomalies networks. Figure 11 shows a large spike in the number of links for the years 1969, 1973, 1974, and 1975 in our positive covariance network. It appears that the early 1970's may have been a period of unusually volatile weather, but more investigation in needed. As of now, we are not sure what caused these spikes but this is an area of possible future investigation.

Figure 12 appears to show a general upward trend in the number of links in our negative covariance network over the years. This suggests that the number of anti-correlated wind cases are increasing with time (i.e. winds are becoming more and more dissimilar between different locations), which might be indicative of some effect of climate change.

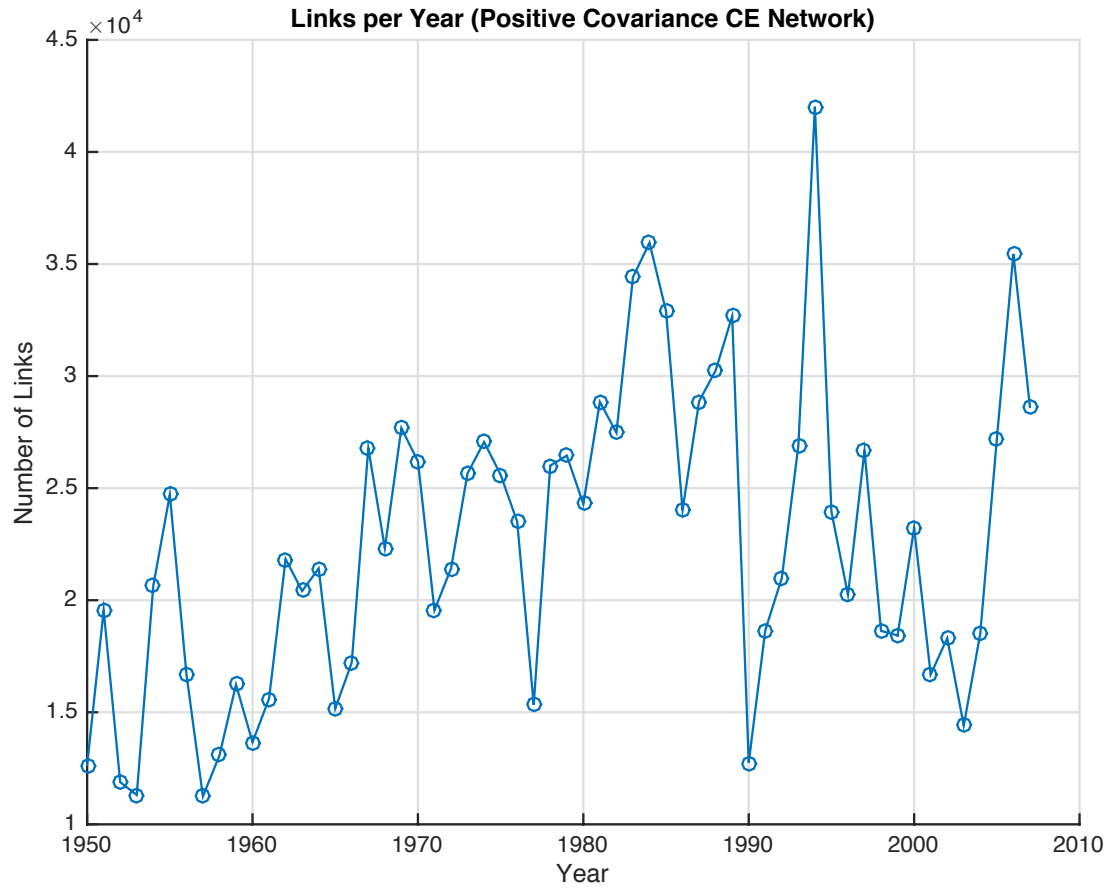

**Figure 9: Links per Year, Positive Covariance Carbon Emissions Network after Thresholding**

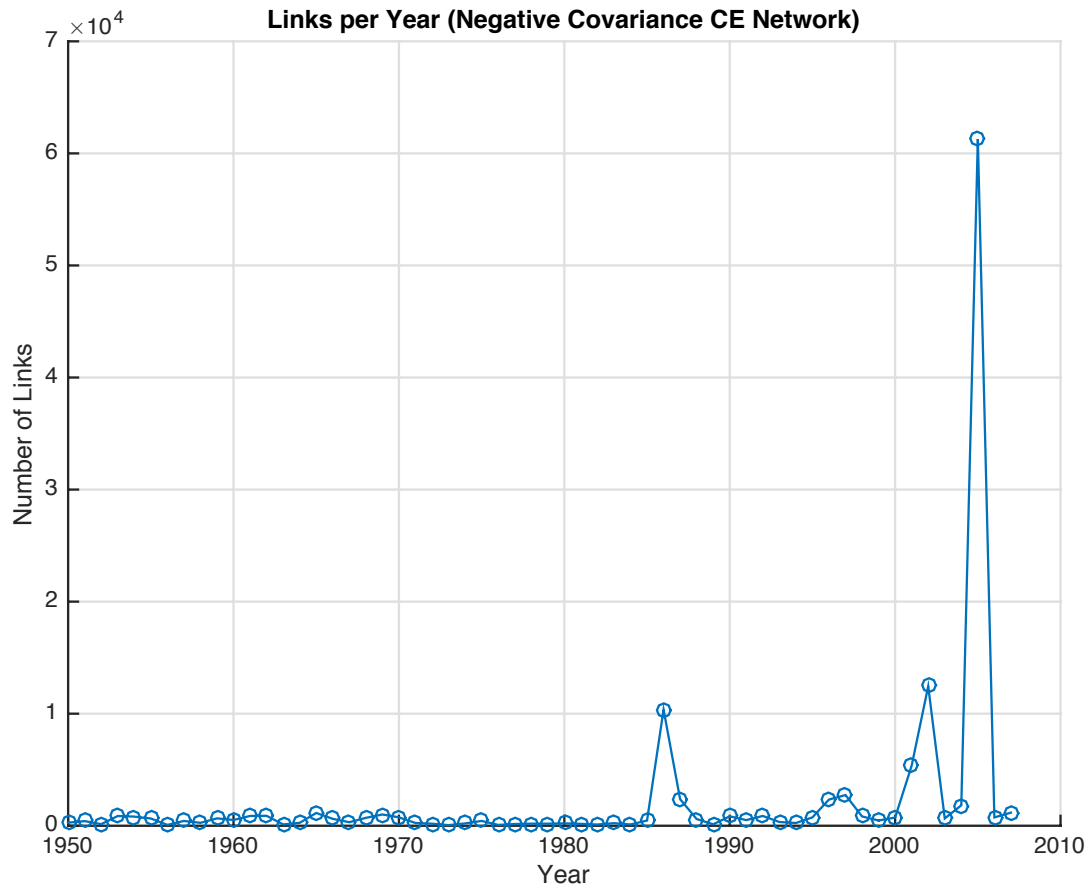

**Figure 10: Links per Year, Negative Covariance Carbon Emissions Network after Thresholding**

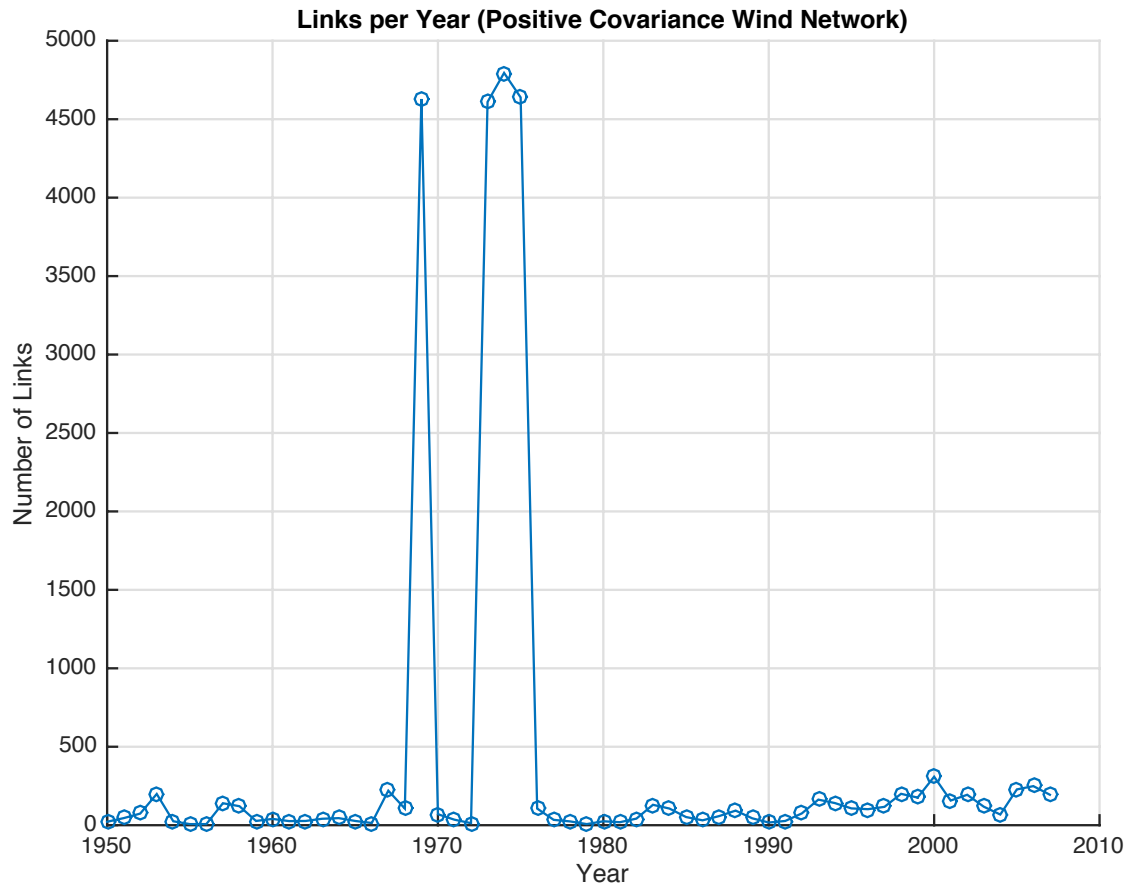

**Figure 11: Links per Year, Positive Covariance Wind Speed Anomalies Network after Thresholding**

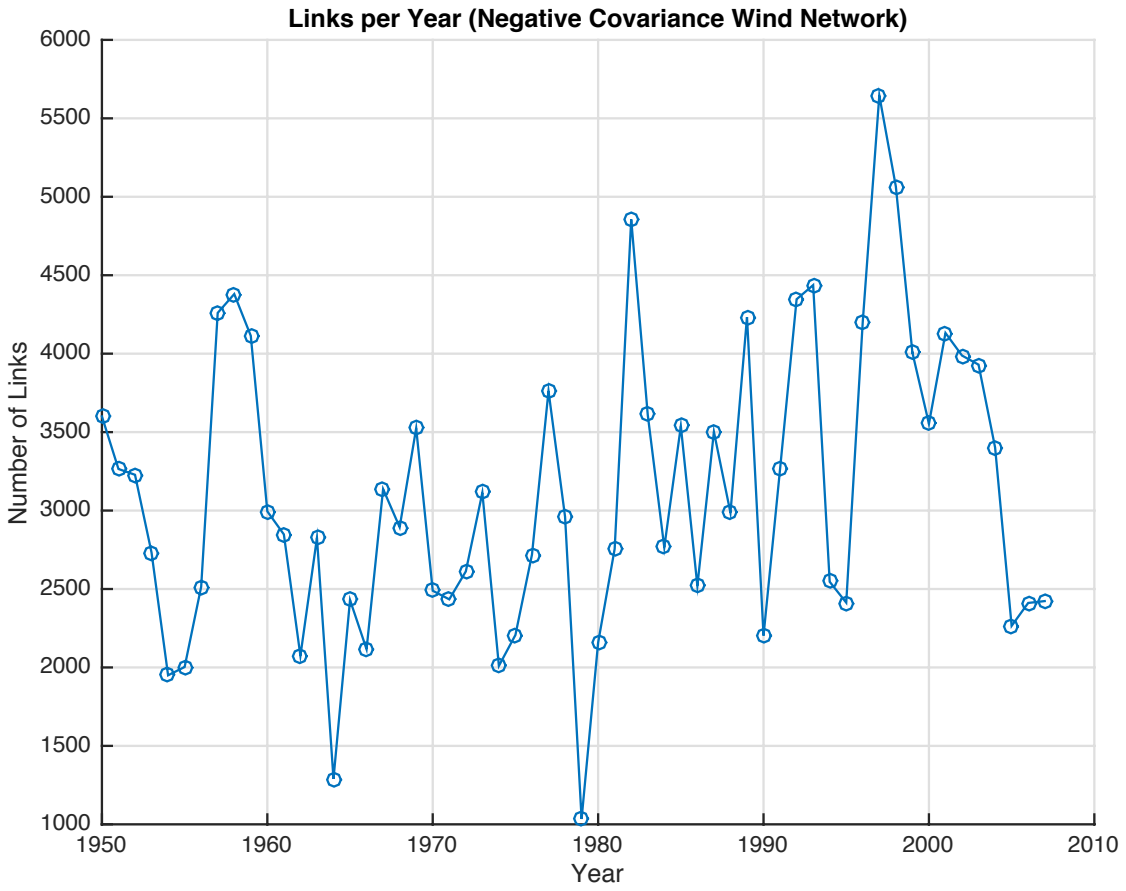

**Figure 12: Links per Year, Negative Covariance Wind Speed Anomalies Network after Thresholding**

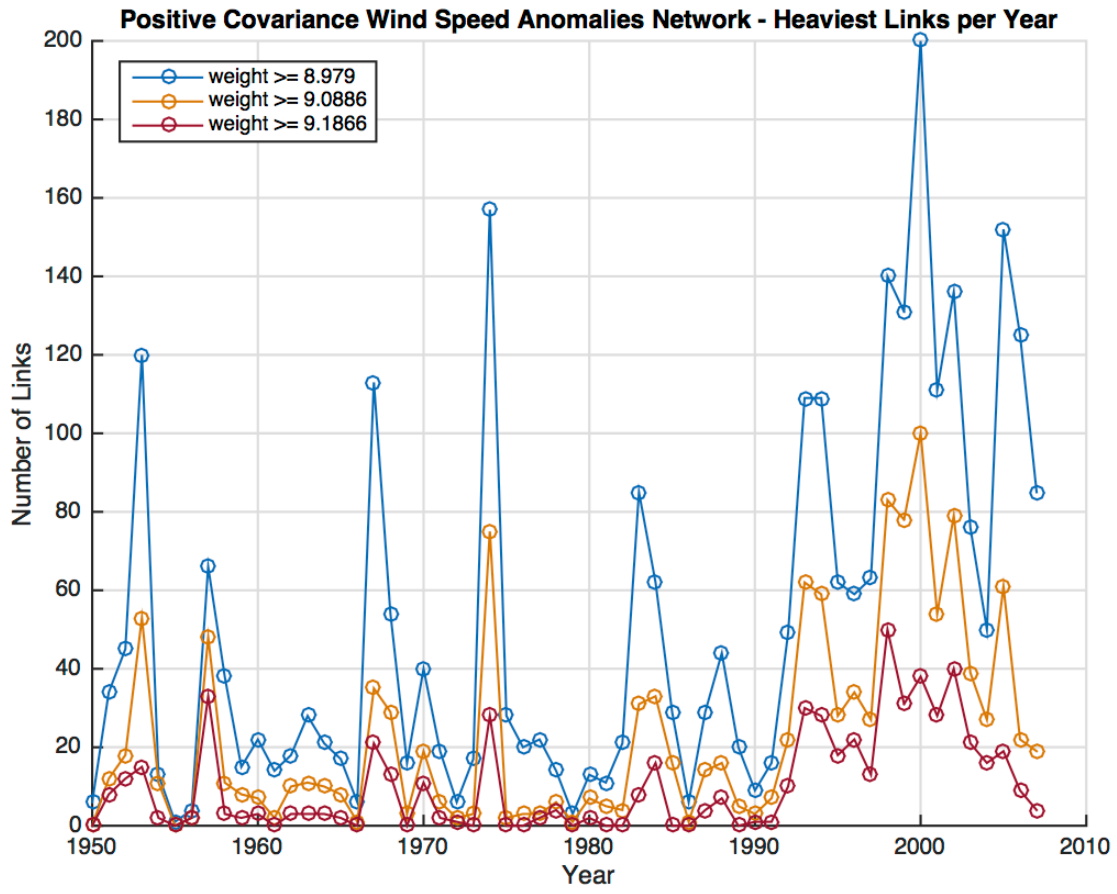

**Figure 13: Heaviest Links per Year, Positive Covariance Wind Speed Anomalies Network**

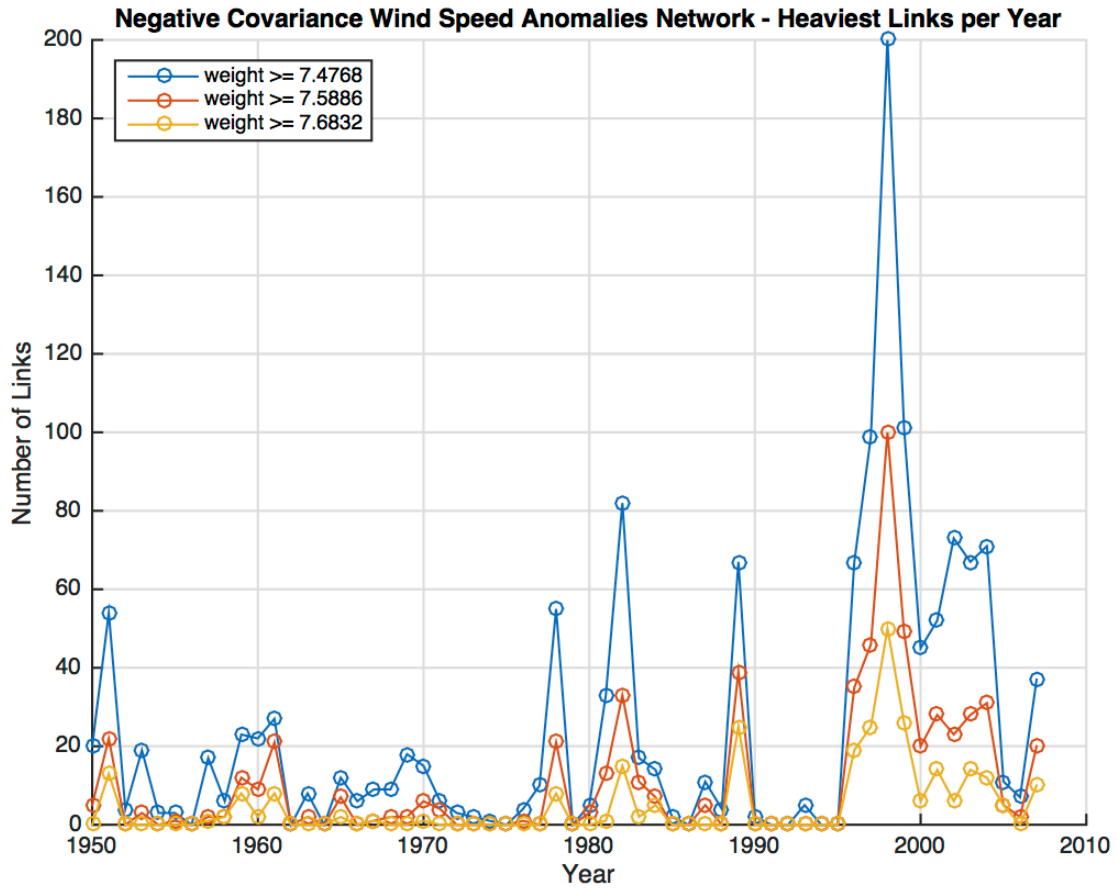

**Figure 14: Heaviest Links per Year, Negative Covariance Wind Speed Anomalies Network**

If we increase the original threshold to focus on the heaviest links this upward trend becomes more significant. The plot in Figure 14 shows the number of links per year above certain thresholds. These thresholds were chose to remove all but the 200, 100, and 50 heaviest links. If we perform the same operation for our positive covariance network we see an upward trend in the heaviest links for this network as well (see Figure 13). This also tells us that the large spikes in Figure 11 are almost completely made up of links whose weight lie between 8.89 and 8.979.

Just as significant is where these links are geographically located. When we mapped the 200 heaviest links from Figure 13 (our positive covariance network for wind speed anomalies) on a globe, we found that the links occurred exclusively over the southern end of the Indian Ocean (a known corridor for tropical cyclones) and Antarctica.

When we did the same operation for links from Figure 14 (our negative covariance network for wind speed anomalies), the links occurred in a tighter cluster over the southern end of the Indian Ocean alone. See Figure 15 and 16 for the geographic link maps for our wind speed anomalies networks for the year 1998. In addition, when compared to the link from our positive covariance network, the links from our negative covariance network are long range links that connect non-adjacent parts of the globe.

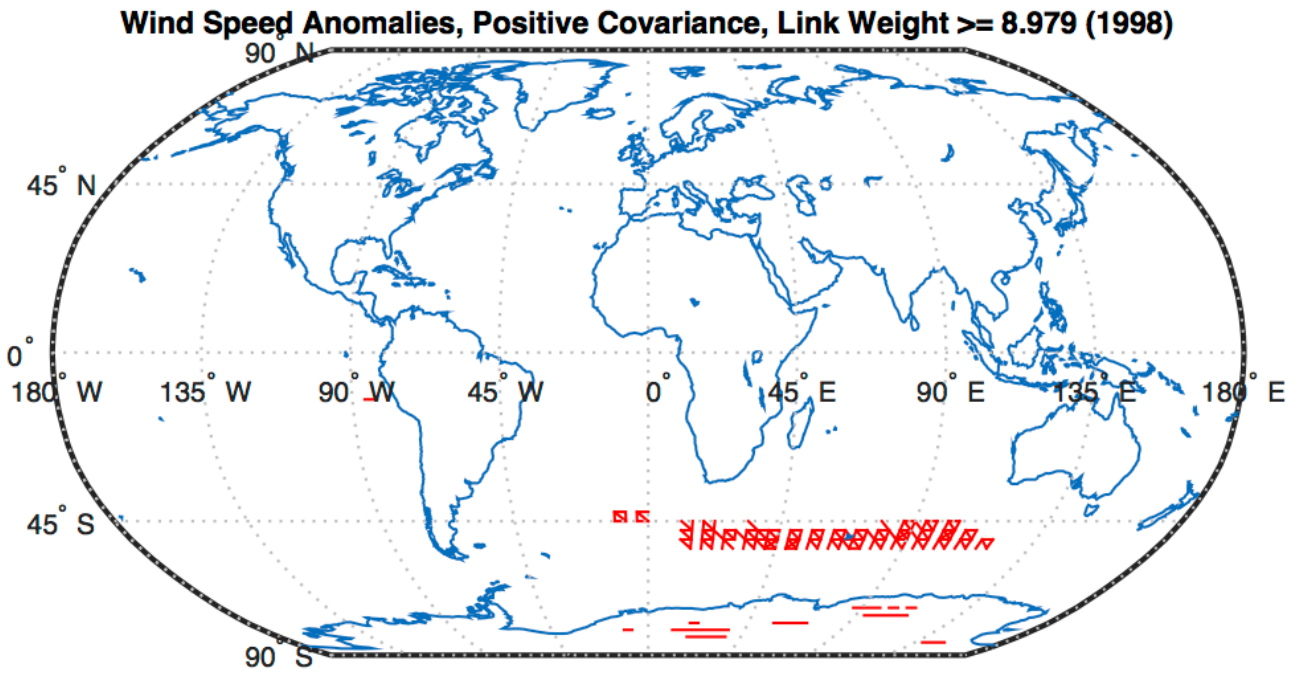

**Figure 15: Map of Wind Speed Anomalies Network with Positive Covariance, 1998**

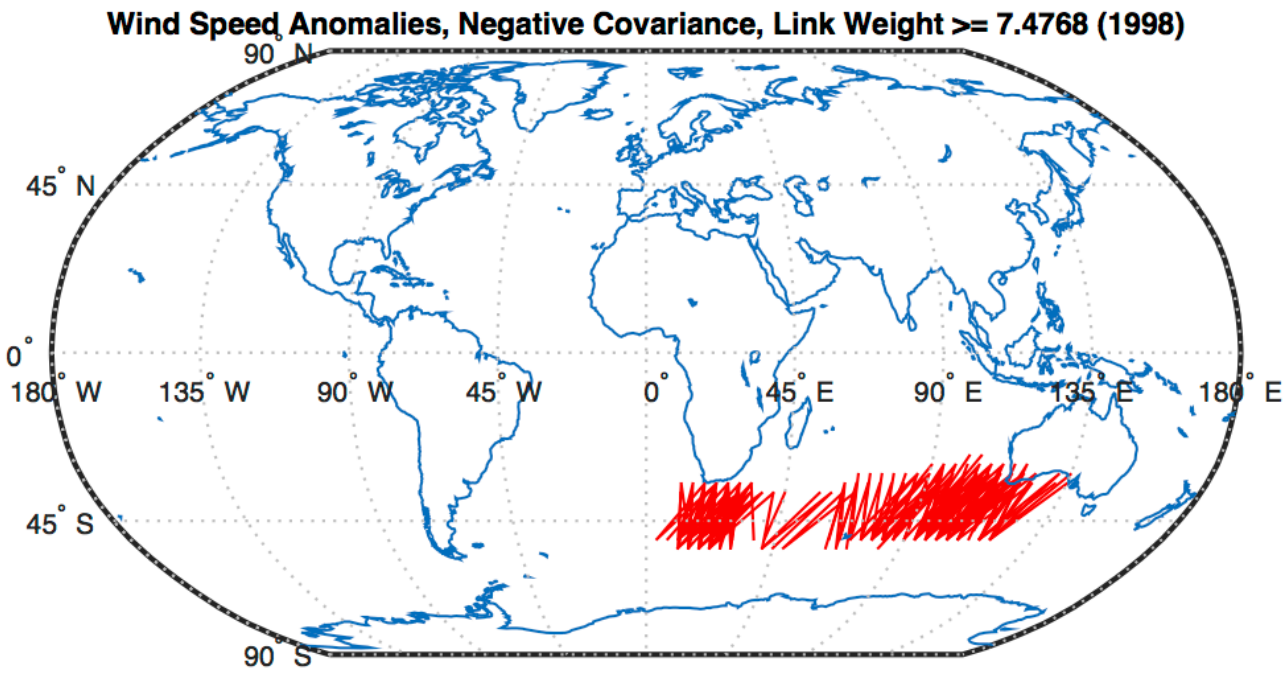

**Figure 16: Map of Wind Speed Anomalies Network with Negative Covariance, 1998**

After seeing this result, since these links are based on the maximum and minimum cross-covariance values between the node they connect, we investigated the time delay of the links that yielded the link weights. Figure 17 shows the distribution of link time delays for the Negative Covariance Wind Speed Anomalies Network. All links in the network had a time delay between -1 days and 2 days. When we focused on the time delay of the 200 heaviest links in the network (Figure 18), the range reduced to 0 days or 1 day. Links with a zero-day time delay (the largest single category of links in both plots) imply that the anomalous conditions at one end of the link are not caused by the conditions at the other end, but rather that an external event outside of our network might be the cause of the anomalous conditions that appear at both ends of the link at the same time.

Another avenue of investigation was looking at the correlation between the number of links in the wind anomalies networks and the number of weather disturbance recorded in the Southwest Indian ocean, a known corridor for tropical cyclones. Using data from [9], Figure 19 shows the number of storms recorded in the Southwest Indian Ocean. The blue line shows the number of storms per year with a rating of Tropical Depression or greater given by MFR (Météo-France office based on the Réunion island). The orange line shows the number of storms per year with a rating of Intense Tropical Depression or greater given by MFR. From the plot, we can see a clear increasing trend in the number of storms per year in the same geographic region as the 200 heaviest links in our positive and negative covariance wind anomalies networks. Computing a correlation value between the number of storms great than or equal to Tropical Depression and the number of links in our networks reveals a weak positive correlation: 0.4761 for positive covariance links and 0.3679 for negative covariance links. When considering the correlation between the number of storms great than or equal to Intense Tropical Cyclone and number of positive covariance links, the highest correlation value found between the data sets is achieved at 0.5457 (a potentially significant positive correlation).

## 7. CONCLUSION & FUTURE WORK

The southwest region of the Indian Ocean is a known corridor for tropical cyclones. The appearance of the heaviest links in our wind speed anomalies networks over this area and their increasing trend over time suggests that our network has captured a real, physical phenomenon which might be indicative of some effect of climate change. The positive correlation between the increasing trend of the heaviest links in the positive covariance network and the increasing number of intense tropical cyclones in the Southwest Indian Ocean suggests a connection between the two.

More investigation is still required Figures 10 and 11 show strange spikes in the number of links of our networks that we cannot yet explain. Correlations between the southwest region of the Indian Ocean and the other metrics (temperature anomalies and carbon emissions) are still unexplored. As stated in Section 4, future work could explore these correlations by performing community detection on a multi-layer network, where each layer of the network is one of the metric networks, in the hope that we find communities of nodes that span across the metric networks. This would signify areas of the globe that experience anomalous behavior at the same time across different metrics.

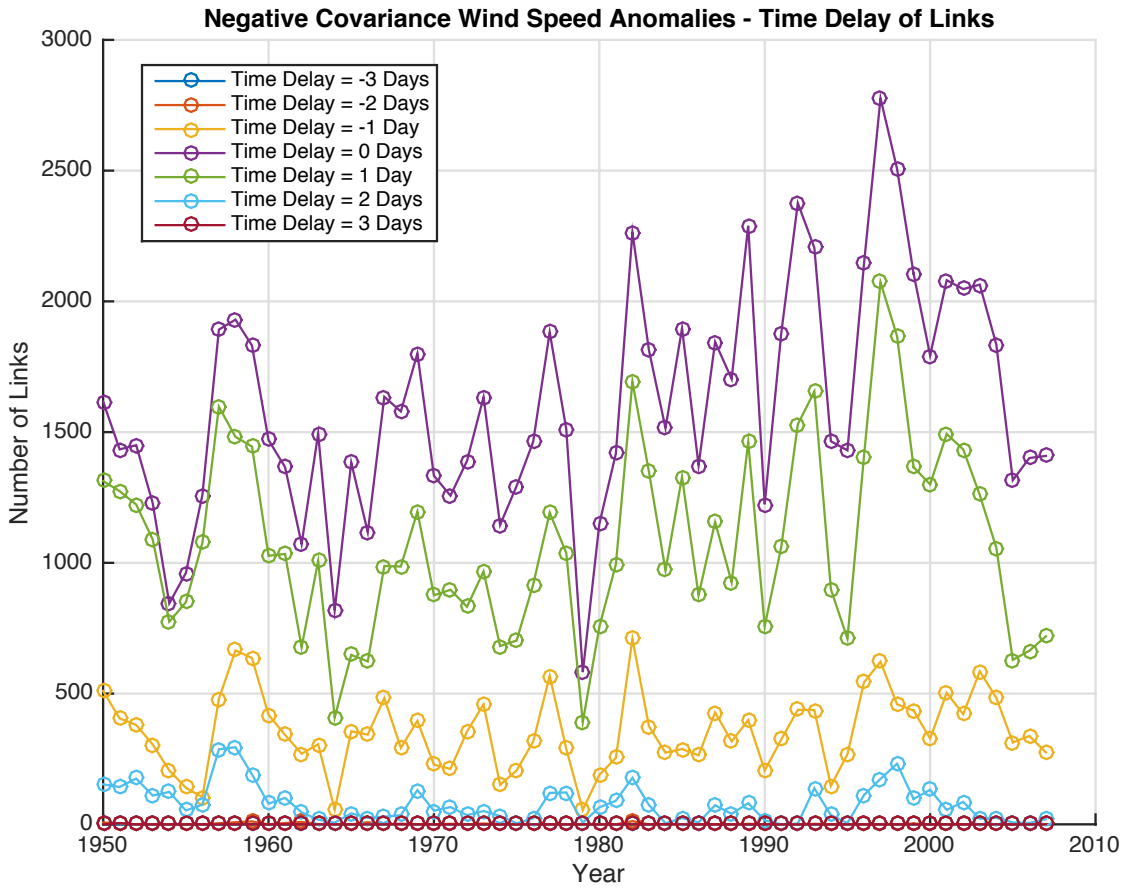

**Figure 17: Distribution of Link Time Delay, Negative Covariance Wind Speed Anomalies Network**

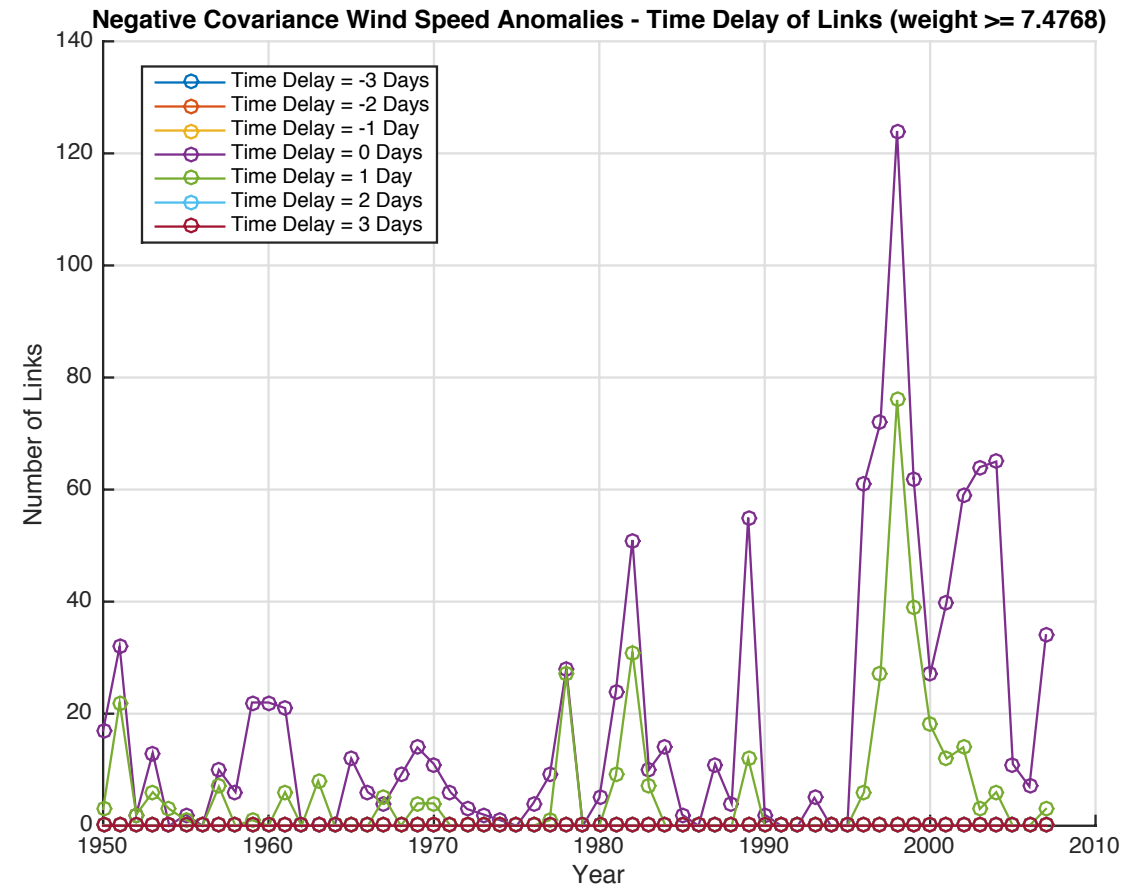

**Figure 18: Distribution of Time Delay for 200 Heaviest Links, Negative Covariance Wind Speed Anomalies Network**

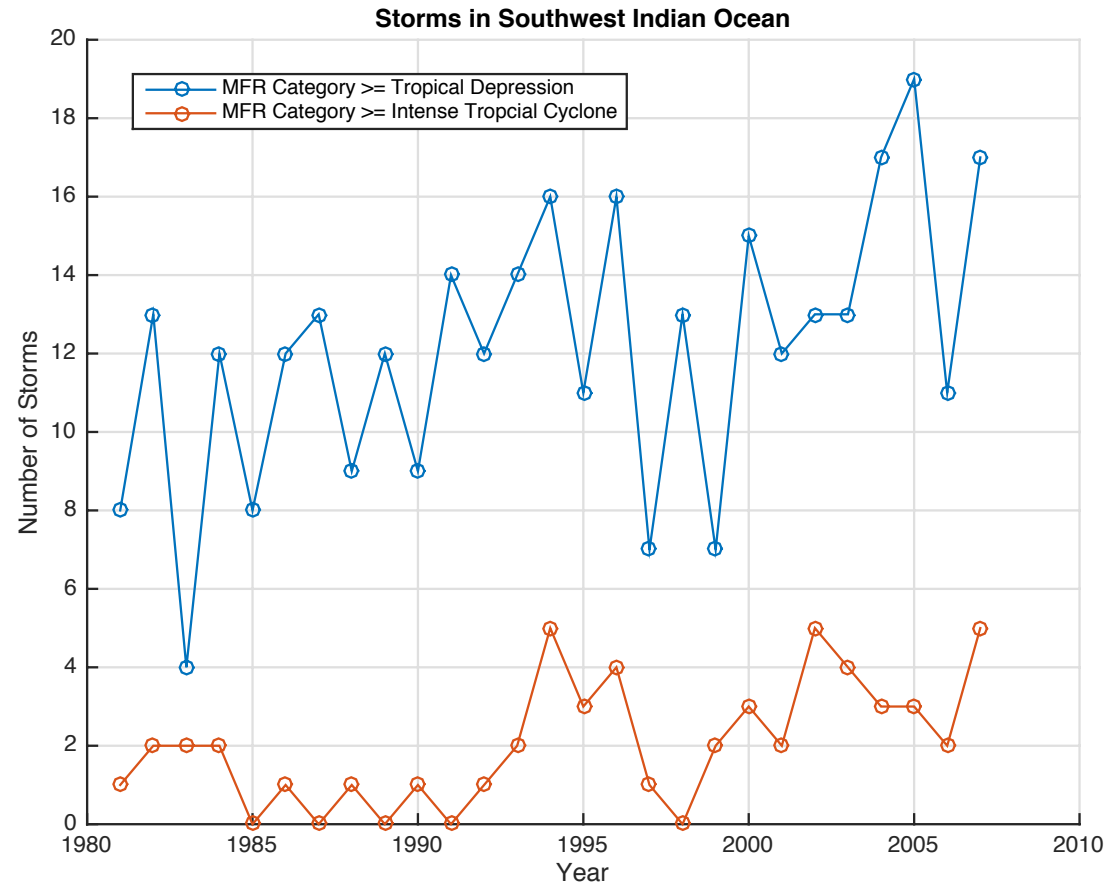

**Figure 19: Number of Storms in Southwest Indian Ocean**

Finally, this project aimed to look at global carbon emissions, global temperature anomalies, and global wind pattern anomalies. Through a modified approach based on [4], we are beginning to see significant links in the wind anomalies in the southwest region of the Indian Ocean. The hope is that this project will shed some light on how man-made carbon emissions are affecting global wind patterns.

## References

[1] J. Caesar, "HadGHCND - gridded daily temperatures," Met Office Hadley Centre, 10 February 2015. [Online]. Available: http://www.metoffice.gov.uk/hadobs/hadghcnd/. [Accessed 22 September 2016].

[2] T. G. M. a. R. A. Boden, "Global, Regional, and National Fossil-Fuel CO2 Emissions," U.S. Department of Energy, 22 09 2016. [Online]. Available: http://cdiac.ornl.gov/CO2_Emission/gridded. [Accessed 22 09 2016].

[3] B. Dattore, "NCEP Climate Forecast System Reanalysis (CFSR) 6-hourly Products, January 1979 to December 2010," 11 August 2016. [Online]. Available: http://rda.ucar.edu/datasets/ds093.0/. [Accessed 22 September 2016].

[4] A. G. Y. B. Y. W. a. S. H. O. Guez, "Global climate network evolves with North Atlantic Oscillation phases: Coupling to Southern Pacific Ocean," EPL, vol. 103, no. 68006, pp. 68006-p1-68006-p5, 14 October 2013.

[5] M. Moyer, "Climate Change May Mean Slower Winds," Nature America, Inc, 1 October 2009. [Online]. Available: http://www.scientificamerican.com/article/climate-change-ay-mean-slower-winds/. [Accessed 22 September 2016].

[6] B. Oskin, "Watch Carbon Pollution Spread Across the Planet," Live Science, 18 November 2014. [Online]. Available: http://www.livescience.com/48798-carbon-dioxide-global-computer-model.html#ooid=05NHZycTqiSuUIE3KIded0oHhfByfvQA. [Accessed 22 September 2016].

[7] S. Eichelberger, J. McCaa, B. Nijssen and A. Wood, "Climate Change Effects On Wind Speed," North American Windpower, pp. 0-0, 01 July 2008.

[8] S. C. Pryor, R. J. Barthelmie, D. T. Young, E. S. Takle, R. W. Arritt, D. Flory, W. J. Gutowski Jr., A. Nunes and J. Roads, "Wind speed trends over the contiguous United States," Journal of Geophysical Research, vol. 114, no. D14, pp. 0-0, 23 July 2009.

[9] Unknown, "South-West Indian Ocean tropical cyclone," Wikipedia.org, 23 July 2016. [Online]. Available: https://en.wikipedia.org/wiki/South-West_Indian_Ocean_tropical_cyclone. [Accessed 03 12 2016].